\documentclass[twocolumn,superscriptaddress,amsmath,amssymb]{revtex4}

\usepackage{graphicx}
\usepackage{dcolumn}
\usepackage{bm}
\usepackage{amsmath}

\begin{document}

\title{Work function seen with sub-meV precision through laser photoemission}

\author{Y.~Ishida}
\email{ishiday@issp.u-tokyo.ac.jp}
\affiliation{Center for Correlated Electron Systems, Institute for Basic Science, Seoul 08826, Republic of Korea}
\affiliation{ISSP, The University of Tokyo, Kashiwa, Chiba 277-8581, Japan}

\author{J.~K.~Jung}
\affiliation{Center for Correlated Electron Systems, Institute for Basic Science, Seoul 08826, Republic of Korea}
\affiliation{Department of Physics and Astronomy, Seoul National University, Seoul 08826, Republic of Korea}

\author{M.~S.~Kim}
\affiliation{Center for Correlated Electron Systems, Institute for Basic Science, Seoul 08826, Republic of Korea}
\affiliation{Department of Physics and Astronomy, Seoul National University, Seoul 08826, Republic of Korea}

\author{J.~Kwon}
\affiliation{Center for Correlated Electron Systems, Institute for Basic Science, Seoul 08826, Republic of Korea}
\affiliation{Department of Physics and Astronomy, Seoul National University, Seoul 08826, Republic of Korea}

\author{Y.~S.~Kim}
\affiliation{Center for Correlated Electron Systems, Institute for Basic Science, Seoul 08826, Republic of Korea}
\affiliation{Department of Physics and Astronomy, Seoul National University, Seoul 08826, Republic of Korea}

\author{D.~Chung}
\affiliation{College of Liberal Studies, Seoul National University, Seoul 08826, Republic of Korea}

\author{I.~Song}
\affiliation{Center for Correlated Electron Systems, Institute for Basic Science, Seoul 08826, Republic of Korea}
\affiliation{Department of Physics and Astronomy, Seoul National University, Seoul 08826, Republic of Korea}

\author{C.~Kim}
\affiliation{Center for Correlated Electron Systems, Institute for Basic Science, Seoul 08826, Republic of Korea}
\affiliation{Department of Physics and Astronomy, Seoul National University, Seoul 08826, Republic of Korea}

\author{T.~Otsu}
\affiliation{ISSP, The University of Tokyo, Kashiwa, Chiba 277-8581, Japan}

\author{Y.~Kobayashi}
\affiliation{ISSP, The University of Tokyo, Kashiwa, Chiba 277-8581, Japan}


\begin{abstract}
Electron emission can be utilised to measure the work function of the surface. However, the number of significant digits in the values obtained through thermionic-, field- and photo-emission techniques is typically just two or three. Here, we show that the number can go up to five when angle-resolved photoemission spectroscopy (ARPES) is applied. This owes to the capability of ARPES to detect the slowest photoelectrons that are directed only along the surface normal. By using a laser-based source, we optimised our setup for the slow photoelectrons and resolved the slowest-end cutoff of Au(111) with the sharpness not deteriorated by the bandwidth of light nor by Fermi-Dirac distribution. The work function was leveled within $\pm$0.4~meV at least from 30 to 90~K and the surface aging was discerned as a meV shift of the work function. Our study opens the investigations into the fifth significant digit of the work function.
\end{abstract} 

\maketitle
Among the electronic properties of a crystal surface, the work function ($\phi_{\rm s}$) is one of the fundamentals. It corresponds to the minimum energy required to take out an electron through the surface at 0~K~\cite{CardonaLey,16MaterHor_Kahn}. The values of $\phi_{\rm s}$ serve as a test bench for the theory of surface electronic structures~\cite{36PR_Bardeen,41PR_Smoluchowski,71PRB_LangKohn,03PRB_LDA_Leung} and are relevant to a variety of electronic and chemical phenomena on surfaces and interfaces. The topics include the junction-device behaviours, charge-carrier injection and surface catalytic interactions~\cite{90Nature_Catalytic,10AdvFunct_OrganicInterface,13NAsia_ThinFilmOxideOrganic}.

\begin{figure}[htb]
\begin{center}
\includegraphics[width = 8.9cm]{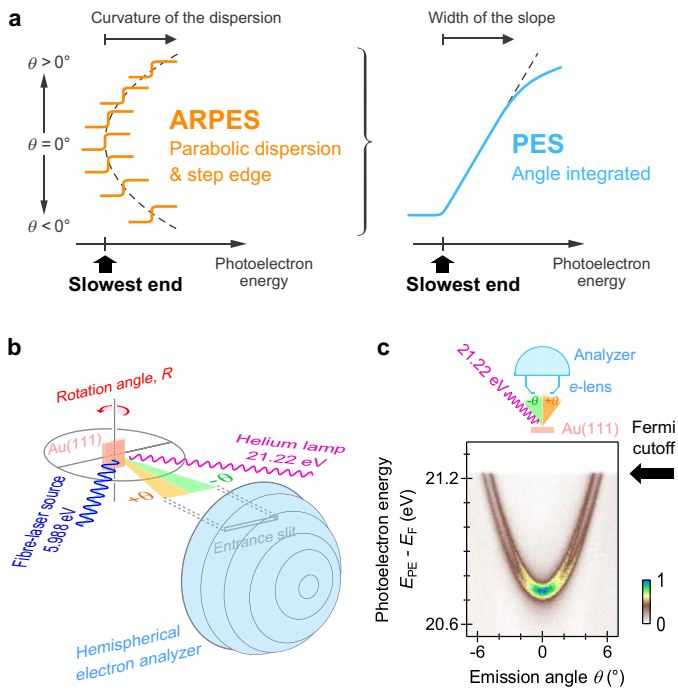}
\caption{\label{fig1}{\bf Detecting the slow photoelectrons.} ({\bf a}) Slow photoelectrons seen via PES and ARPES. When integrated over a certain emission anglular cone about the normal emission ($\theta=0^{\circ}$), the spectrum becomes sloped. ({\bf b}) The ARPES setup equipped with a fibre-laser source and a helium lamp. Au(111) is held at 30~K unless described otherwise. The sample and entrance of the $e$-lens are electrically connected to the common ground. ({\bf c}) Shockley state of sample~1 seen with helium-lamp ARPES.}
\end{center}
\end{figure}

Electrons emitted from a crystal can be utilised to measure $\phi_{\rm s}$ directly because the threshold of the emission is governed by $\phi_{\rm s}$ that acts as a potential barrier on the surface. Thus, thermionic-, field-, and photo-emission techniques have been applied to this end~\cite{CardonaLey}. However, the thresholds are not as sharp as naively expected, and the experimental values of $\phi_{\rm s}$ typically have just two or three significant digits~\cite{CardonaLey}. This number is also typical to other techniques; see Kawano~\cite{08ProgSurfSci_Kawano} and Derry {\it et al.}~\cite{15JVacSciTech_Derry}, in which over 1000 values obtained through various methods are tabulated. The low precision of $\phi_{\rm s}$ has limited in-depth investigations into, for example, its dependence on temperature ($T$)~\cite{49RMP_HerringNichols,86SurfSci_Kiejna}, strain~\cite{07PRB_Cu110_Strain_Sekiba,10EPL_Cu110_Strain_XFWang,12JPhysCond_StrainGraphene,15JPhysCond_StrainTMD,16NCom_Strain_Rubrane}, or even on gravity; see Cardona and Ley~\cite{CardonaLey} and references therein. 

While the primary concern may be the surface contamination, there is also a kinematic effect that can smooth the threshold of the electron emission seen in experiments~\cite{31PR_Fowler,69PR_Spicer_Cu}: Consider an ensemble of electrons each having just the energy to climb the potential barrier $\phi_{\rm s}$; it is only those directed normal to the surface that can indeed climb and be at rest on the outer surface. This illustrates that the number of electrons available on the outer surface diminishes when the threshold is approached. This effect was taken into account by Fowler to explain the rather smooth upturn of the photoyield when the photon energy $h\nu$ was swept across $\phi_{\rm s}$~\cite{31PR_Fowler}, and later, into all other emission techniques~\cite{33PR_DuBridge,06JAP_Jensen_GeneralEEmission}. In the case for ultraviolet and x-ray photoemission spectroscopy (PES), the slow ends of the photoelectron spectra are observed as $\gtrsim$0.1-eV-wide slopes instead of step edges because of the effect~\cite{CardonaLey,69PR_Spicer_Cu}, and the energy level of the slowest photoelectrons ($E_{\rm s}$) has to be read from the point where the slope merges into the background (Fig.~\ref{fig1}a). As a result, the values of $\phi_{\rm s}$ are read typically with two to three significant digits~\cite{96APL_ITO,10ApplSurfSci_Pitfalls,14APL_Akaike} while the number of four does exist in some studies~\cite{12JChemPhys_Yoshinobu_Rh111}. In any case, the degree of certainty is lower than the sub-meV precision routinely attained when setting the energy level of the Fermi cutoff ($E_{\rm f}$) located on the fast end of the spectrum~\cite{12Sci_Okazaki,19CM_HuangZhou_Planck}. 

The kinematic effect upon the emission~\cite{31PR_Fowler,69PR_Spicer_Cu} strongly indicates that, in order to select out the slowest electron among the emitted electrons, not only their energy but also their trajectory has to be resolved. Here, we apply angle-resolved PES (ARPES) to the slow photoelectrons emitted from a crystal surface of a metal. As will be explained in the present study, the slowest end seen in ARPES becomes a step edge (Fig.~\ref{fig1}a) that is intrinsically sharper than the Fermi cutoff. Therefore, ARPES has the potential to locate $E_{\rm s}$ with the degree of certainty much higher than that for locating $E_{\rm f}$. This underlies the precise extraction of the work function via ARPES through the relationship $\phi_{\rm s}=h\nu-(E_{\rm f}-E_{\rm s})$. Simple though it may seem, the precise detection via ARPES was aided by the use of a fibre-laser-based light source~\cite{16RSI_FiberARPES} whose beam can be aligned and focused to take control over the trajectory of the slow photoelectrons in ARPES setups~\cite{14RSI_TiSi}; see the subsequent section, Laser-ARPES setup for slow photoelectrons. The slow end of the photoelectron distribution seen in fibre-laser ARPES retained the sharpness expected in Einstein's theory~\cite{05AnnPhys_Einstein} and allowed us to monitor the work function with sub-meV precision. 
\\

\section*{Results}
\noindent{\bf Laser-ARPES setup for slow photoelectrons}
Slow photoelectrons are vulnerable to fields and their trajectories can easily be bent by the electric and magnetic fields remaining in the spectrometer~\cite{10ApplSurfSci_Pitfalls}. Particularly, the slowest photoelectrons are those that have once been at rest on the outer surface, and taking control of their trajectory from the surface to the electron analyser is the most difficult. As will be explained later, the genuine cutoff of the slowest end seen in ARPES should exhibit the following two features (Fig.~\ref{fig1}a). (1) The slow-side cutoff shows a parabolic angular dispersion. (2) The cutoff is a step edge whose width is not broadened by Fermi-Dirac distribution nor by the bandwidth of light. We set these two features, which was not described in the previous studies of the work function~\cite{CardonaLey,02PRB_Paggel}, as the criteria for the successful ARPES on the slow photoelectrons. 

The technical difficulty to perform reliable ARPES on slow photoelectrons is gradually being removed, and currently, slow photoelectrons of  $\lesssim$1~eV kinetic energy can be passed through modern electron analysers in a controlled trajectory. This advance has partly been propelled by the advent of low-$h\nu$ sources of $\lesssim$10~eV based on lasers~\cite{12Sci_Okazaki,06PRL_Koralek,12RSI_Xe_HarterKyle,14RSI_TiSi,18Rev_Zhou}, wherein the detection of slow photoelectrons becomes a mandatory. To attain a fair angular resolution particularly when the photoelectrons are slow, it is requested to keep the beam diameter less than 0.3~mm at the focal point of the electron lens ($e$-lens) attached to the analyser, and the laser-based sources can easily meet this demand. 

We set up a fibre-laser-based source of $h\nu=5.988$~eV~\cite{16RSI_FiberARPES} and docked it to an ARPES spectrometer equipped with a helium lamp (Fig.~\ref{fig1}b; also see Supplementary Note~1 and Supplementary Fig.~\ref{figS1}). The beam diameter at the focal point of the $e$-lens was set to $\sim$0.1~mm through a procedure that utilises a pin hole~\cite{14RSI_TiSi}. The oscillator of the fibre-laser source was stably mode locked for at least three months, which ensures that the profile of the laser such as the photon energy and band width was locked during the measurements. While fibre-laser ARPES was used to detect the slow photoelectrons, helium-lamp ARPES was also used to characterise the band dispersions and to calibrate the photoelectron energy $E_{\rm PE}-E_{\rm F}$ referenced to the Fermi level of the sample in electrical contact to the analyser; see Methods. 
\\

\noindent{\bf  Preparation of Au(111)} 
As a model system, we investigated the (111) surface of gold. Gold is a noble metal and is an exemplary metal for electrodes. Studies of the work function of gold begun partly in pursuit of a reference standard~\cite{66APL_Mercury_Huber,66APL_Mercury_Riviere}, and the values for Au(111) have been found to be in 5.2\,-\,5.6~eV~\cite{08ProgSurfSci_Kawano,15JVacSciTech_Derry,14APL_Akaike,82SurfSci_SARPES}. It is well known that the work function is sensitive to the surface quality at the atomic level, and there is consensus that the cleaner the surface is, the higher the work function especially for materials with the values greater than $\sim$4~eV~\cite{70PRB_Eastman}. We prepared two samples of Au(111) (samples~1 and 2) through cycles of Ar-ion bombardment and annealing (Methods). The dispersion of the Shockley surface state formed on Au(111)~\cite{96PRL_LaShell,04PRB_Hoesch,12PRB_BYKim} was observed by using helium-lamp ARPES (Fig.~\ref{fig1}c). The band dispersion for sample~2 was slightly sharper than that for sample~1 (Supplementary Fig.~\ref{figS2} and Supplementary Note~2). As we shall see later, sample~2 indeed exhibited higher $\phi_{\rm s}$ than that of sample~1. 
\\

\begin{figure}[htb]
\begin{center}
\includegraphics{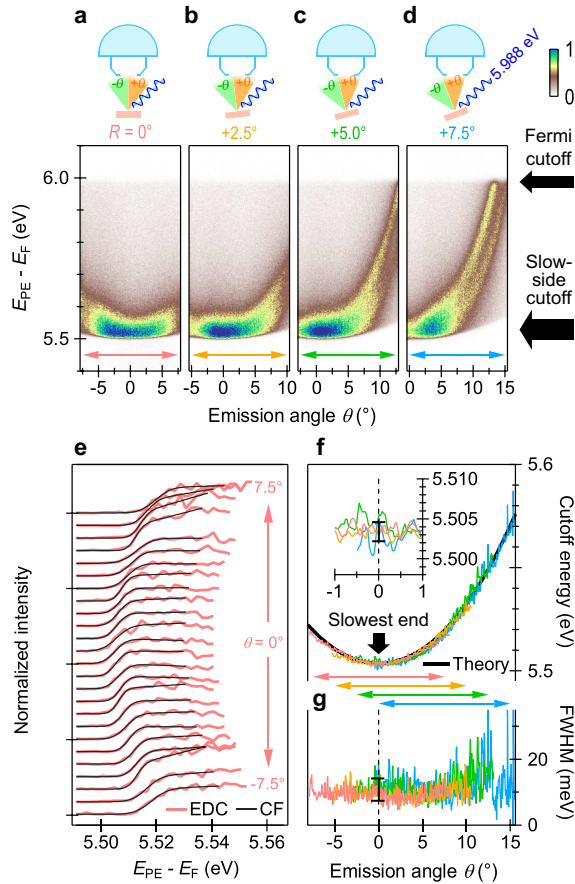}
\caption{\label{fig2}{\bf Sharp and parabolic cutoff on the slow side.} ({\bf a-d}) Photoelectron distributions detected with fibre-laser ARPES. The sample was rotated from $R=0$ ({\bf a}) to $7.5^{\circ}$ ({\bf d}). ({\bf e}) EDCs and CFs around the cutoff acquired in the $R=0^{\circ}$ configuration. ({\bf f} and {\bf g}) Cutoff energy ({\bf f}) and width ({\bf g}) of the CFs as functions of $\theta$. The inset to {\bf f} shows the cutoff energy around $\theta=0^{\circ}$, with the bar indicating $5.5034\pm0.0012$~eV for the values in [-0.5, 0.5$^{\circ}$]. The bar in ({\bf g}) indicates $10.8\pm3.4$~meV for the values in [-7.5, 15$^{\circ}$].}
\end{center}
\end{figure}

\noindent{\bf Slow end seen with fibre-laser ARPES}
The panels in Fig.~\ref{fig2}a-d show the photoelectron distributions obtained when Au(111) of sample~1 was illuminated by the fibre-laser source. The fastest photoelectrons formed the Fermi cutoff at the known photon energy 5.988~eV of the source, which ensures the accuracy of $E_{\rm PE}-E_{\rm F}$ calibrated by using the He I$\alpha$ line. As the sample was rotated from $R=0$ (Fig.~\ref{fig2}a; normal-emission configuration) to 7.5$^{\circ}$ (Fig.~\ref{fig2}d), the Shockley surface band became fully visible up to the Fermi cutoff. However, the bottom of the band was truncated by the cutoff on the slow side because the electrons excited from the bottom could not overcome the work function. The visibility of the Shockley band down to the cutoff provides a credit that our ARPES setup was operating agreeably for the slowest photoelectrons even without applying any bias voltage on the sample; also see Supplementary Fig.~\ref{figS2}.  

Our main observations for Fig.~\ref{fig2}a-d are of the slow side cutoff. First, the cutoff energy depended on the emission angle of the photoelectrons ($\theta$) and exhibited a parabolic dispersion bottomed at the surface normal, $\theta=0^{\circ}$. Second, the slow-side cutoff was a sharp step edge in contrast to the $\gtrsim$0.1-eV-wide slopes observed in ultraviolet and x-ray PES~\cite{96APL_ITO,10ApplSurfSci_Pitfalls,14APL_Akaike,12JChemPhys_Yoshinobu_Rh111}. The parabolic cutoff is distinct from the parabolic boundary that occurs in the photoelectron distribution mapped onto energy-momentum ($E$\,-\,$k$) plane. As is well known, the mapping function~\cite{18RSI_Mapping} has a range that sets a parabolic boundary in $E$\,-\,$k$ plane whose bottom occurs not in the surface normal but in the direction towards the $e$-lens. Such a boundary in $E$\,-\,$k$ plane can be seen for example in Bovet {\it et al.}~\cite{04PRL_SecondaryARPES} The cutoff herein is observed in $E$\,-\,$\theta$ plane and its bottom followed the surface normal when the sample was rotated from $R=0$ to $7.5^{\circ}$. Therefore, the parabolic cutoff is intrinsic to the emission from the surface. 

In order to quantify the sharp-and-parabolic cutoff, we performed a fitting procedure and extracted the energy and width of the cutoff as functions of $\theta$. In the procedure, we first set a cutoff function (CF), which was a step distribution function of a linear slope convoluted with a Gaussian; for the explicit  form of CF, see Methods. Then, with the CF, we fitted the energy distribution curves (EDCs) of the data shown in Fig.~\ref{fig2}a-d. The case for the $R=0^{\circ}$ distribution is presented in Fig.~\ref{fig2}e. The four parabolic curves nicely overlapped each other (Fig.~\ref{fig2}f) and the full width at half maximum of the Gaussian (FWHM: $\gamma$) was averaged as $\gamma=10.8\pm3.4$~meV (Fig.~\ref{fig2}g). The cutoff energy around $\theta=\pm0.5^{\circ}$, in which there are 115 data points, was averaged as 5.5034~eV with one standard deviation of $\sigma=1.2$~meV (inset to Fig.~\ref{fig2}f). The energy at the bottom of the parabolic cutoff will subsequently be identified as the absolute value of the work function. Note, the overlap of the four dispersions seen in Fig.~\ref{fig2}f shows that the dispersion shifted with the known interval of 2.5$^{\circ}$ along the emission angle axis and ensures the angular scaling of $\theta$. 

With the sharp cutoff as a measure, we were able to discern the aging of the surface. After keeping Au(111) of sample~1 in the spectrometer for 10~hours, the sample surface became less clean and the cutoff shifted downwards; see, Supplementary Fig.~\ref{figS3} and Supplementary Note~3. The shift was as small as 5.5~meV and was attributed to the reduction in the work function due to residual gas weakly physisorbed on the surface. The reduction is consistent with the trend that the work function lowers as the surface becomes less clean~\cite{70PRB_Eastman}. There was no discernible shift of the Fermi cutoff during the 10~hours, which indicated that the analyser condition was stable during the measurement; see Methods. 

\begin{figure}[htb]
\begin{center}
\includegraphics[width = 8.9cm]{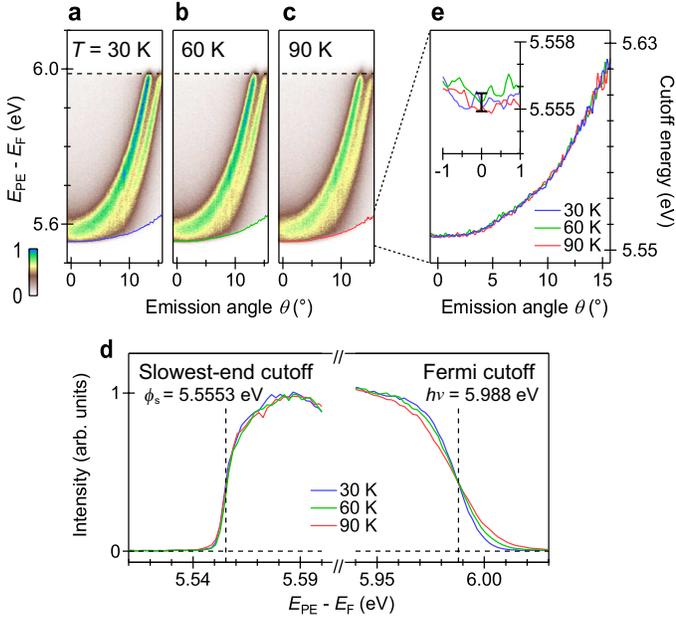}
\caption{\label{fig3}{\bf Temperature dependence of the cutoff.} ({\bf a-c}) Photoelectron distributions recorded at the temperatures of $30$ ({\bf a}), 60 ({\bf b}) and 90~K ({\bf c}) on sample~2. ({\bf d}) Temperature dependence of the distribution curves across the slowest-end cutoff at $\theta=0^{\circ}$ and across the Fermi cutoff. ({\bf e}) The slow-side cutoffs at various temperatures. Inset shows the expanded view around the emission angle $\theta=0^{\circ}$, in which one standard deviation $\sigma$ of 0.4~meV for the values in [-0.5, 0.5$^{\circ}$] is indicated by an error bar.}
\end{center}
\end{figure}

In a separate run of the measurement (sample~2), we varied the temperature from 30 to 90~K and monitored the cutoff (Fig.~\ref{fig3}a-c). In contrast to the Fermi cutoff, the slow-end cutoff remained sharp (Fig.~\ref{fig3}d). The width of the slowest end around $\theta=\pm0.5^{\circ}$, in which there are 45 data points, was $\gamma=8.3\pm1.0$~meV and the cutoff energy of the slowest end stayed at 5.5553~eV with one standard deviation of $\sigma=0.4$~meV (Fig.~\ref{fig3}e). That is to say, there was no temperature dependence in the work function with the precision of $\pm$0.4~meV/60~K $=\pm0.08k_{\rm B}$, where $k_{\rm B}$ is the Boltzmann constant. The absolute value $5.5553$~eV was higher than the value $5.5034$~eV for sample~1 and is comparable to the highest reported work function of $5.6\pm0.1$~eV on Au(111) obtained through a Fowler plot~\cite{82SurfSci_SARPES}. Thus, with the work function as the measure~\cite{70PRB_Eastman}, the surface quality of sample~2 was better than that of sample~1 and was comparable to that studied in Pescia {\it et al.}~\cite{82SurfSci_SARPES}

Before explaining why the cutoff on the slow side appears sharp and parabolic, we point out that the cutoff is not only truncating the Shockley band but also the background signal, as clearly seen at $\theta>10^{\circ}$ in Fig.~\ref{fig2}d. The background signal could originate from bulk bands as well as photoelectrons that have encountered some scattering~\cite{03JES_Strocov}. This observation indicates that, when explaining the features of the cutoff, the photoelectrons forming the Shockley band and background signal have to be treated on equal footings. We thus consider a model for whatever photoelectrons that pass through a homogeneous surface characterised by a single work function $\phi_{\rm s}$. 
If there had been multiple edges in the spectrum~\cite{10ApplSurfSci_Pitfalls}, the surface would have been judged as non-uniform, or patchy~\cite{10ApplSurfSci_Pitfalls,49RMP_HerringNichols}.
\\

\noindent{\bf  Trajectory of the threshold photoelectrons} 
The sharp-and-parabolic appearance of the cutoff can be understood by considering the trajectory of ``threshold photoelectrons." Below, we first explain the photoelectron refraction, or the kinematic constraints upon the emission across the surface~\cite{31PR_Fowler,69PR_Spicer_Cu,CardonaLey}, and define the threshold photoelectrons. Then, we consider their trajectories from the surface to the entrance of the $e$-lens that collects the photoelectrons into the hemispherical analyser. 

\begin{figure}[htb]
\begin{center}
\includegraphics[width = 8.9cm]{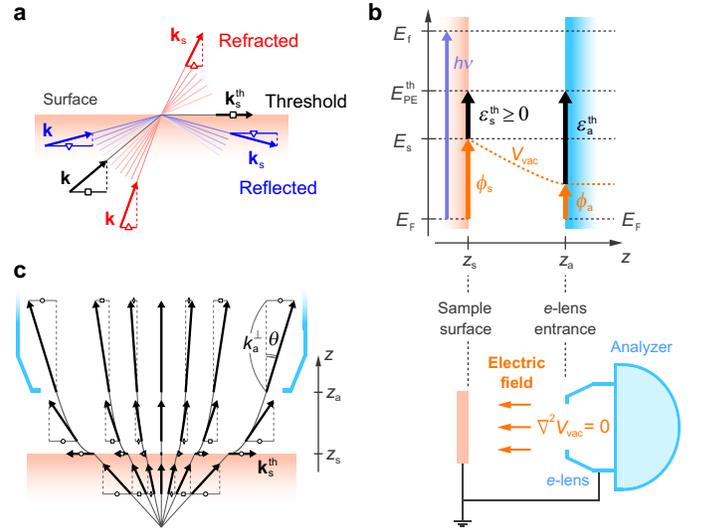}
\caption{\label{fig4}{\bf Trajectory of the threshold photoelectrons.} ({\bf a}) The refraction and reflection of the photoelectrons and the definition of the threshold photoelectrons. ({\bf b}) The energy diagram for the threshold photoelectrons. The lower section shows the electric field existing between the sample and $e$-lens separated with the working distance of $z_{\rm a}-z_{\rm s}\sim32$~mm (Supplementary Note~1). The vacuum level $V_{\rm vac}({\mathbf r})$ is the solution to the Poisson equation. ({\bf c}) The trajectory of the threshold photoelectrons dragged by the electric field. The threshold photoelectron emitted normal to the surface has the smallest momentum (shortest black arrow) when entering the $e$-lens, and hence, is the slowest and forms the slowest-end cutoff. $\theta$ is the emission angle seen by the analyser. }
\end{center}
\end{figure}

When passing through the surface, photoelectrons are refracted because the work function acts as a potential barrier. As shown in Fig.~\ref{fig4}a, the angle of refraction becomes large as the angle of incidence increases, and at a critical angle, the photoelectrons travel tangential to surface. We call these tangential photoelectrons the threshold photoelectrons. Their kinetic energy on the surface is $\varepsilon_{\rm s}^{\rm th}\,=\,(\hbar{\mathbf k}_{\rm s}^{\rm th})^2/2m\,\geq\,0$, where $m$ is the electron mass and $\hbar{\mathbf k}_{\rm s}^{\rm th}$ is the momentum that is parallel to the surface by definition. The energy level of a threshold photoelectron $E_{\rm PE}^{\rm th}$ can be described as (Fig.~\ref{fig4}b),
\begin{equation}\label{eq1}
E_{\rm PE}^{\rm th}=E_{\rm F}+\phi_{\rm s}+\varepsilon_{\rm s}^{\rm th}\geq E_{\rm F}+\phi_{\rm s}\equiv V_{\rm vac}^{z_{\rm s}},   
\end{equation}
where $V_{\rm vac}^{z_{\rm s}}$ is the vacuum level just outside the surface.

The threshold photoelectrons cannot reach the $e$-lens as long as they are traveling tangential to the surface. Here, we are reminded that, even when the sample and $e$-lens are electrically connected, electric fields can exist between the two; see Fig.~\ref{fig4}b. The vacuum level is a solution to the Poisson equation, $\nabla^2V_{\rm vac}({\mathbf r})=0$, with the boundary condition set by the work function on the vacuum boundary. Thus, $V_{\rm vac}$ differs by $\varDelta\phi=\phi_{\rm s}-\phi_{\rm a}$ between the sample and entrance of the $e$-lens, where $\phi_{\rm a}$ is the work function of the material that coats the interior of the $e$-lens and analyser. When $\varDelta\phi>0$, the threshold photoelectrons can take off the surface and be dragged towards the $e$-lens. Their kinetic energy at the $e$-lens entrance becomes $\varepsilon_{\rm a}^{\rm th}=\varepsilon_{\rm s}^{\rm th}+\varDelta\phi$ (Fig.~\ref{fig4}b). For the case when $\varDelta\phi<0$, see later. 

Analytic solutions for the trajectory exist when we can regard the electric field to be directed along the surface normal ($z$) (Fig.~\ref{fig4}c). This arrangement is similar to an infinitely large parallel-plate capacitor, but each plate is made of different materials. Then, while dragged, the momentum parallel to the surface is unchanged. At the $e$-lens entrance, the momentum along $z$ (${\hbar}k_{\rm a}^{\perp}$) can be obtained through $({\hbar}k_{\rm a}^{\perp})^2/2m={\varDelta}\phi$, and the nominal emission angle ($\theta$) seen by the analyser becomes $\tan\theta=|{\mathbf k}_{\rm s}^{\rm th}|/k_{\rm a}^{\perp}$ (Fig.~\ref{fig4}c). Thus, equation~\eqref{eq1} can be described as
\begin{equation}\label{eq2}
 E_{\rm PE}^{\rm th}-E_{\rm F}=\phi_{\rm s}+\varDelta\phi \tan^2\theta. 
\end{equation}
Equation~\eqref{eq2} illustrates that, when entering the $e$-lens, the energy of the threshold photoelectron exhibits a parabolic angular dispersion, and this is the dispersion detected by the analyser. In Fig.~\ref{fig2}f, we overlaid the curve of equation~\eqref{eq2} with $\varDelta\phi=0.9$~eV, which determines the curvature, and the bottom of the dispersion $\phi_{\rm s}=5.5034\pm0.0012$~eV is identified as the absolute value of the work function. 

For completeness, we present the dispersion of the angle-resolved cutoff when the sample is applied with negative bias voltage $-v/e$ with respect to the analyser: 
\begin{equation}\label{eq3}
E_{\rm PE}^{\rm th}-E_{\rm F}=\phi_{\rm s}+v+(\varDelta\phi+v)\tan^2\theta. 
\end{equation}
Here, $E_{\rm F}+v$ becomes the Fermi level of the sample. When $\varDelta\phi+v>0$, the threshold photoelectrons are dragged towards the $e$-lens and the intrinsic cutoff becomes visible. With increasing $v$, the dispersion shifts upwards in energy and its curvature (prefactor of $\tan^2\theta$) is increased. The increase of the curvature is in accordance to the photoelectron acceptance cone being tunable with $v$~\cite{15UltraMic_Tusche,19RSI_Yamane}. At $v\gtrsim25$~eV, the lowering of the work function due to the Schottky effect~\cite{49RMP_HerringNichols} can exceed 1~meV and may prevail when seen with the sub-meV precision; see Methods. When integrated over a certain anglular cone about the normal emission, the cutoff is smoothed into a slope (Fig.~\ref{fig1}a right schematic) as seen in ultraviolet and x-ray PES~\cite{96APL_ITO,10ApplSurfSci_Pitfalls,14APL_Akaike,12JChemPhys_Yoshinobu_Rh111}. This slope seen in the normal-emission configuration is the lineshape formulated in Cardona and Ley~\cite{CardonaLey} and in Krolikowski and Spicer~\cite{69PR_Spicer_Cu} that took the Fowler effect into account, although the role of $\varDelta\phi$ was not explicated. According to equation~\eqref{eq3}, the width of the slope in the integrated spectrum becomes wider as $v$ is increased, while the angle-resolved cutoff remains sharp; see Fig.~\ref{fig1}a for the relationship between the curvature seen in ARPES and width of the sloped region seen in PES. 

The demonstration of the sub-meV precision measurement under the biased condition is presented in Supplementary Fig.~\ref{figS4}. We performed the measurement at room temperature and at the pressure of $6\times10^{-10}$~Torr on the surface of an exfoliated highly-oriented pyrolitic graphite (HOPG)~\cite{16RSI_FiberARPES}; the width of the slowest-end cutoff was as narrow as $\gamma=8.0$~meV when one battery of  $v/e=1.62$~V was attached, and the energy of the slowest end around $\theta=\pm0.5^{\circ}$ was leveled within one standard deviation of $\sigma=0.15$~meV when the number of the attached batteries was varied from one to four. For the details of the demonstration, see Supplementary Note~4. 
\\

\noindent{\bf On the role of the monochromaticity} 
It was shown in the previous section that the parabolic dispersion of the slow-end cutoff depends solely on how the threshold photoelectrons were dragged from the outer surface to the $e$-lens; it does not depend on how the threshold photoelectrons were generated. Whatever the value of $h\nu$ may be, the threshold photoelectrons emerged on the outer surface line up on the identical dispersion when entering the $e$-lens. This point is implicit in equations~\eqref{eq1}\,-\,\eqref{eq3} as well as the definition of the work function $\phi_{\rm s}=E_{\rm s}-E_{\rm F}$ being independent of $h\nu$. Thus, the cutoff on the slow end is not blurred by the bandwidth of light ($\gamma_{h\nu}$) besides not being affected by the temperature dependence of the Fermi-Dirac function. Therefore, the cutoff can be observed with the resolution ($\gamma_{\rm a}$) set by the analyser and the stability of the bias voltage. This point is in strong contrast to the bands and Fermi cutoffs seen with the convoluted resolution $(\gamma_{\rm a}^2+\gamma_{h\nu}^2)^{1/2}$. The width of the slow-end cutoff being slightly larger for sample~1 ($10.8\pm3.4$~meV) than that for sample~2 ($8.3\pm1.0$~meV) can be attributed to the degree of inhomogeneity of the work function ($\gamma_{\phi}$) within the probed area set by the $\sim$0.1-mm beam size. That is to say, the width of the cutoff seen in ARPES is $\gamma=(\gamma_{\rm a}^2+\gamma_{\phi}^2)^{1/2}$. When $\gamma_{\phi}\to0$, the width $\gamma$ becomes the direct measure of $\gamma_{\rm a}$. 

The only but important role for the light to be monochromatic was to precisely locate the energy level of the Fermi cutoff $E_{\rm f}$, which was the requisite to refer to the absolute value of the work function~\cite{96APL_ITO,CardonaLey}; see Methods. If the purpose was only to observe the parabolic cutoff and monitor the relative value of the work function, the monochromaticity of the source is not needed; whatever sources that can generate an ensemble of excited electrons in the crystal, or an ensemble of threshold photoelectrons traveling along the outer surface, can be utilised. For example, synchrotron light can be used~\cite{02PRB_Paggel} even when its photon energy is drifting; intense femtosecond infrared pulses that can generate muti-photon photoelectrons can be used~\cite{95JCP_Aeschlimann,06PRL_MultiPPE} provided that the intense field of the pulse does not significantly alter the work function~\cite{49RMP_HerringNichols}; deuterium lamps and electron guns, the latter in the setup of momentum-resolved electron-energy loss spectroscopy~\cite{15RSI_EELS,18Science_MEELS}, may also be used if the beam size can be reduced sufficiently. 
\\

\section*{Discussion}
In summary, we applied 6-eV fibre-laser ARPES to a model system Au(111) and investigated the trajectory of the slow photoelectrons that marginally overcame the work function. The slow end of the photoelectron distribution was successfully detected. It showed a parabolic angular distribution bottomed at normal emission. Moreover, the cutoff was a step edge and was sharper than the Fermi cutoff. A kinematic model described the sharp-and-parabolic cutoff as follows: The bottom of the dispersion is the cutoff formed by the slowest photoelectrons; the curvature of the dispersion depends on the slope of the vacuum level across the sample and $e$-lens; the cutoff is a step edge whose sharpness is indifferent to the bandwidth of light. Thereby, we derived the work function with one standard deviation as small as $\sigma=0.4$~meV and demonstrated the potential of ARPES that is intrinsically free from the Fowler effect.  

Because the work function is sensitive to the surface quality at the atomic level~\cite{CardonaLey,41PR_Smoluchowski,49RMP_HerringNichols,70PRB_Eastman}, it is practically impossible to obtain the {\it accurate} value of $\phi_{\rm s}$ of an ideal crystal surface that has no contamination. In this sense, the present contribution just adds two more values to the literature, namely $\phi_{\rm s}=5.5034\pm0.0012$ and $5.5553\pm0.0004$~eV for the studied surfaces of Au(111). However, the significance is in their {\it precision}~\cite{08ProgSurfSci_Kawano,15JVacSciTech_Derry}. This owes to the capability of ARPES to resolve the direction of the emission, which was the prerequisite to observe the spectral cutoff formed only by the slowest photoelectrons emitted along the surface normal. We demonstrated that the temperature dependence is as small as $d\phi_{\rm s}/dT<\pm0.08k_{\rm B}$ (Fig.~\ref{fig3}), which puts strong constraints on the theory that typically predicts $d\phi_{\rm s}/dT=\mathcal{O}(k_{\rm B})$~\cite{49RMP_HerringNichols,86SurfSci_Kiejna}, and suggests that there can be a compensation between the bulk and surface contributions to $d\phi_{\rm s}/dT$ for Au(111)~\cite{86SurfSci_Kiejna}. The precision also allowed us to discern the surface aging over 10~hours as the 5.5-meV shift of the work function (Supplementary Fig.~\ref{figS3}). The precise measurement by using ARPES can open new opportunities to monitor the work function on the surface, for example, in ambient conditions~\cite{03PRL_Atomistic,10RSI_AmbientARPES,17APE_Takagi_Ambient,19RSI_Ambient_Pillow}, under controlled application of strain~\cite{18PRB_Pfau_Detwin,19NPJ_StrainSRO}, during phase transitions or crossovers~\cite{92PRL_YBCO_TDep,98JES_HF_Schonhense,18PRL_MoTe} and upon irradiation of intense femtosecond pulses~\cite{11Science_Fausti,14PRB_Kaiser,12LaserRev_Bovensiepen,14RSI_TiSi,17PRL_Bi2212_Miller}.  
\\

\section*{Methods}
\noindent{\bf ARPES setup on Au(111)}
The Au(111) surface of $\sim$1\,$\times$\,1~cm$^2$ was prepared by repeating 1.8-keV Ar-ion sputtering and 550-K annealing on a single crystal of gold. The temperature on the surface during the annealing was monitored by a pyrometer whose emissivity was set to 0.05. The pressure during the final 10-min annealing of samples 1 and 2 were less than $7\times10^{-10}$ and $4\times10^{-10}$~Torr, respectively. After the annealing, the sample was cooled at a rate of $\sim$15~K/min down to 400~K and then transferred to an ARPES chamber at a pressure of $3.7\times10^{-11}$~Torr. A homemade light source based on optical fibres doped with ytterbium (Yb)~\cite{16RSI_FiberARPES,08OptExp_Kobayashi,17OptExpress_Nakamura} was docked to the chamber equipped with a hemispherical analyser (Scienta-Omicron, DA30-L), a retractable helium lamp, a 6-axis manipulator, and a temperature controlled cryostat. The analyser had a one-dimensional entrance slit as illustrated in Fig. 1a, and photoelectrons directed into the slit were detected at once. The ARPES dataset was thus obtained as a two-dimensional matrix spanned by energy and emission angle. The wattage of the 5.988-eV probe light was $\sim$1~$\mu$W. For details of the fibre-laser source, see Supplementary Note~1. The temperature of the sample was controlled in the range of 30 to 90~K. The degassing from the cryostat was below the detection limit of an ion gauge, whose read of the pressure did not vary during the temperature variation; see Supplementary Fig.~\ref{figS3}e.
\\

\noindent{\bf Energy reference for extracting the work function}
As is well established (for example, see Park {\it et al.}~\cite{96APL_ITO}), the absolute value of the work function $\phi_{\rm s}$ can be obtained by using photoemission spectroscopy, which owes to the working principle of the hemispherical analyser, or electrostatic deflection analysers in general, when used in a constant pass-energy mode~\cite{CardonaLey}. Then, the photoelectron energy level ($E_{\rm PE}$) can be referenced to $E_{\rm F}$ of the sample in electrical contact to the analyser without any other inputs except the photon energy ($h\nu$) of the source~\cite{96APL_ITO}. In the present study, the spectrum was recorded by sweeping the retardation voltage ($v_{\rm r}/e$) applied in the $e$-lens section while setting the pass energy of the hemispherical analyser to 2~eV. The photoelectron energy $\varepsilon\equiv E_{\rm PE}-E_{\rm F}$ was calibrated as follows: We performed helium-lamp ARPES on gold evaporated on a sample holder at $T=10$~K, and the value of the retardation voltage $v_{\rm r}^{\rm F}/e$ at which the photoelectron excited from $E_{\rm F}$ passed through the hemisphere was determined from the Fermi cutoff, as routinely performed in ARPES studies~\cite{12Sci_Okazaki,19CM_HuangZhou_Planck}; then, the reference of $\varepsilon$ was taken so that the cutoff appeared at the photon energy of the He I$\alpha$ line (21.2180~eV). Note, whatever the sample's work function may be, the photoelectrons excited from $E_{\rm F}$ pass through the analyser when $v_{\rm r}/e$ matches the identified $v_{\rm r}^{\rm F}/e$, provided that the condition of the analyser such as the work function $\phi_{\rm a}$ of the material that coats the interior of the $e$-lens and analyser is not changed during the measurement. When the spectra are displayed as functions of $\varepsilon$, the cutoff energy of the slowest end becomes the absolute value of the work function, whereas the Fermi cutoff appears at the photon energy of the source. We reiterate here that there is no need to know the absolute value of the analyser's work function $\phi_{\rm a}$, only its stability, for calibrating $\varepsilon$ and for reading the absolute value of the sample's work function $\phi_{\rm s}$ from the spectrum.
\\

\noindent{\bf Fitting procedure for the cutoffs}
The cutoff energy $\varepsilon_{\rm s}^{\rm th}$ and width of the cutoff on the slow side were extracted by fitting the EDC at each emission angle with the step-edge-type cutoff function CF: $\int S(\varepsilon')\varTheta(\varepsilon'-\varepsilon_{\rm s}^{\rm th})G(\varepsilon-\varepsilon'; \gamma)d\varepsilon'$. Here, $\varTheta(\varepsilon)$ is a step function which is unity when $\varepsilon>0$ but is zero otherwise, $G(\varepsilon; \gamma)$ is a Gaussian, $\gamma$ is the FWHM of the Gaussian, and $S(\varepsilon)$ is the spectrum to be cut off and is represented by a linear slope, or the Taylor expansion up to the first order about the cutoff energy $\varepsilon_{\rm s}^{\rm th}$.
\\

\noindent{\bf Estimation of the Schottky effect}
All the spectra were taken with the sample and analyser electrically connected to common ground. Therefore, the lowering of the work function due to the Schottky effect $\varDelta\phi_{\rm s}=-(e^3E/4\pi\varepsilon_0)^{1/2}$ (see Fig.~1 of Herring and Nichols~\cite{49RMP_HerringNichols}) was at most caused by the residual electric field $E\sim\varDelta\phi/e(z_{\rm a}-z_{\rm s})$ (Fig.~\ref{fig4}b). Here, $\varepsilon_0$ is the vacuum permittivity and $z_{\rm a}-z_{\rm s}\sim32$~mm is the working distance between the sample surface and entrance of the $e$-lens (Supplementary Note~1). Inserting $\varDelta\phi=0.9$~eV obtained from the curvature of the parabolic cutoff (main text), $\varDelta\phi_{\rm s}\sim0.2$~meV, which is smaller than the standard deviation of the found values of $\phi_{\rm s}$ (main text). When a negative bias voltage $-v/e$ is applied to the sample, then $E\sim(\varDelta\phi+v)/e(z_{\rm a}-z_{\rm s})$, and $\varDelta\phi_{\rm s}$ can exceed 1~meV at $v\gtrsim25$~eV.  
\\

\noindent{\bf Code availability} 
Code used for the fitting procedure is available from the corresponding author on reasonable request.
\\

\noindent{\bf Data availability}
The authors declare that the data supporting the findings of this study are included within the paper and available from the corresponding author on reasonable request.
\\

\noindent{\bf Acknowledgements} 
Y.I., T.O.\ and Y.K.\ acknowledge T.~Nakamura for standardising the so-called $N$-box designed for compacting and stabilising the Yb-doped fibre-laser oscillator. This work was conducted under the ISSP-CCES Collaborative Programme and was supported by the Institute for Basic Science in Republic of Korea (Grant Numbers IBS-R009-Y2 and IBS-R009-G2) and by JSPS KAKENHI (Grant Numbers 17K18749, 18H01148, 19K22140 and 19KK0350). Y.I.\ acknowledges the financial support by the University of Tokyo for the sabbatical stay at Seoul National University.
\\

\noindent{\bf Author contributions}
Y.I.\ conceived the project; Y.I.\ and T.O.\ set up the fibre-laser system under the supervision of Y.K.; J.K.J., M.S.K., J.K., Y.S.K.\ and I.S.\ maintained the ARPES system constructed at Seoul National University under the direction of C.K.; Y.I.\ prepared the sample with instructions from J.K.J.\ and I.S.; Y.I.\ performed ARPES measurements with support from J.K.J., M.S.K., J.K, Y.S.K.\ and D.C.; Y.I.\ analysed the data and wrote the manuscript with input from all authors.
\\

\noindent{\bf Competing interests}
The authors declare no competing financial interests.
\\

\noindent{\bf Materials \& correspondence}
\noindent Materials and correspondence should be addressed to Y.I.: ishiday@issp.u-tokyo.ac.jp.

\section*{Supplementary Notes}

\renewcommand{\figurename}{Supplementary Fig}
\renewcommand{\thefigure}{S\arabic{figure}}
\def\thesection{\arabic{section}}
\setcounter{figure}{0}

\subsection*{Supplementary Note~1: The fibre-laser source} 
Here, we sketch the fibre-laser system and describe some unique points when it was used for performing ARPES on the slow photoelectrons. The system was originally developed at ISSP, University of Tokyo as a compact light source for 95-MHz time-resolved ARPES~\cite{16RSI_FiberARPES}. It was reconstructed at Seoul National University and was upgraded, as described below, to operate stably in an ordinary laboratory environment of moderate temperature stability, vibration level, and air cleanliness. 

\begin{figure}[htb]
\begin{center}
\includegraphics{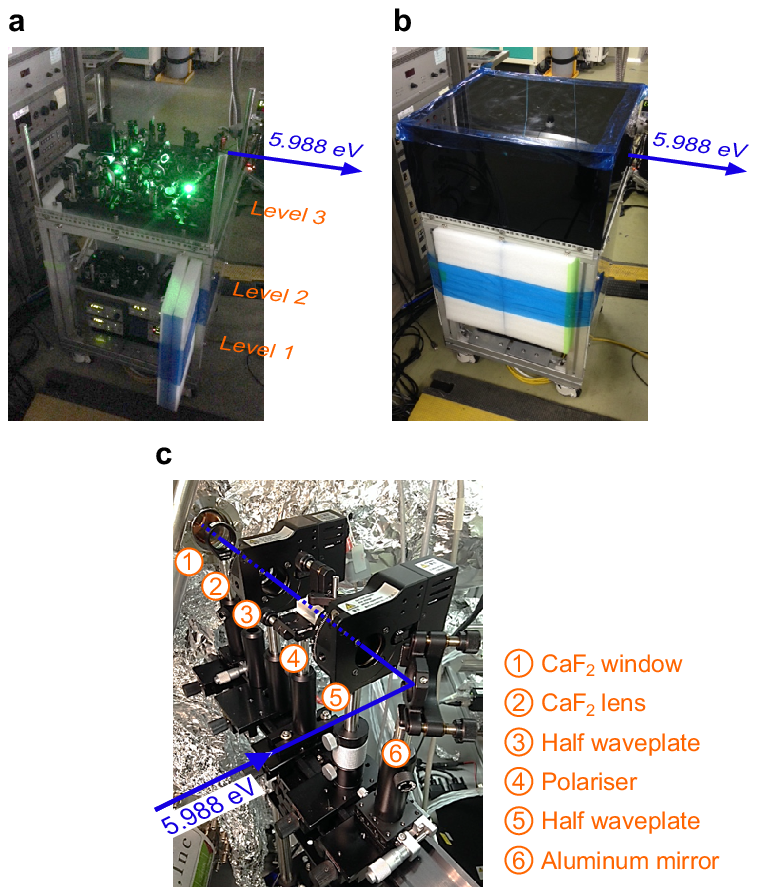}
\caption{\label{figS1}{\bf The fibre-laser system.} The container of the system without ({\bf a}) and with a cover ({\bf b}). {\bf c} The beam line directing the 5.988-eV laser harmonics into the ultrahigh-vacuum chamber of the spectrometer through a CaF$_2$ window. For the devices\,$\textcircled{\scriptsize1}$\,-\,$\textcircled{\scriptsize6}$, see Supplementary Note~1.}
\end{center}
\end{figure}

\begin{figure*}[htb]
\begin{center}
\includegraphics{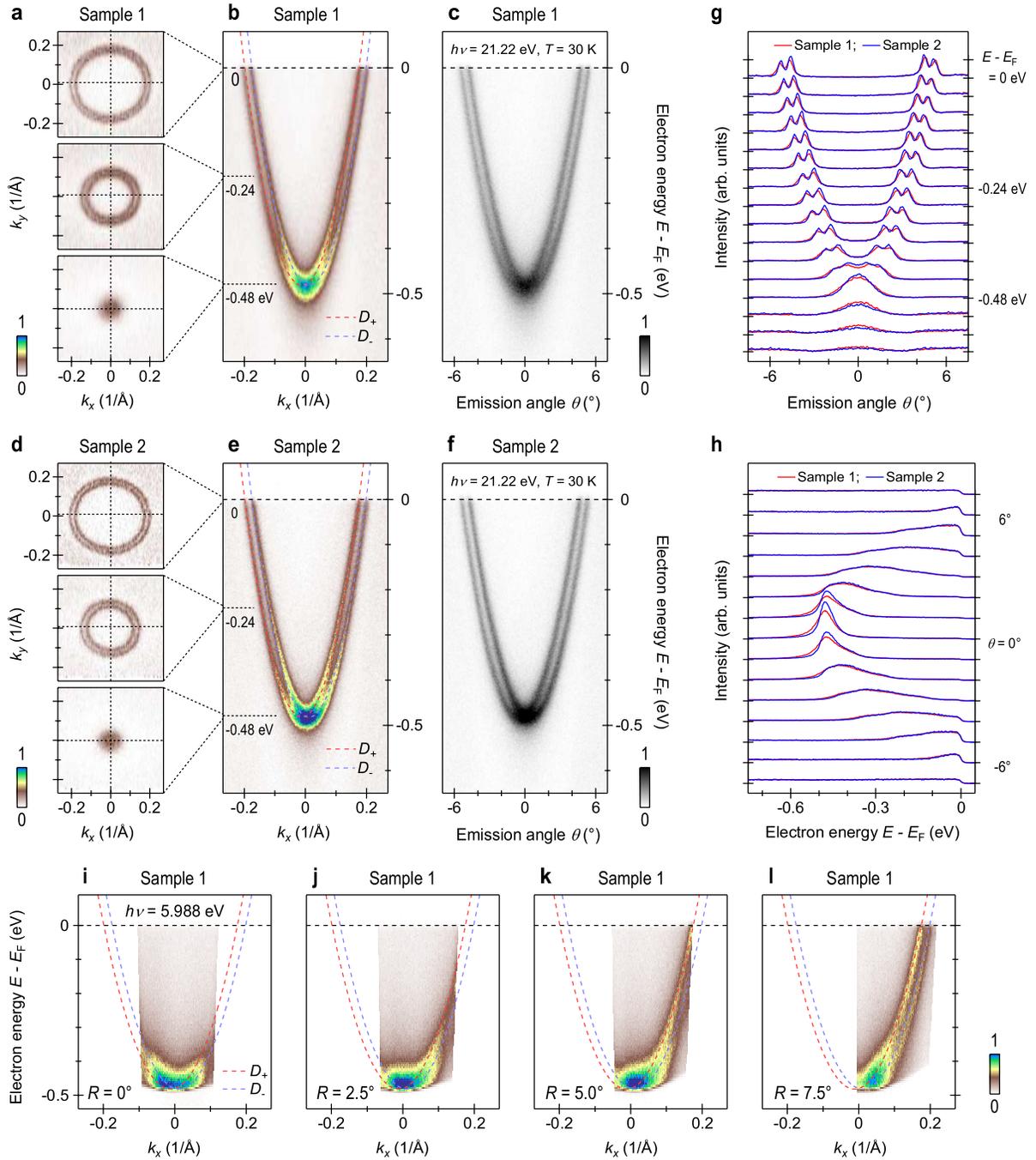}
\caption{\label{figS2}{\bf The surface bands of Au(111).} ({\bf a-c}) ARPES results for sample~1 acquired by using a helium lamp. ({\bf d-f}) Results for sample~2. Shown in {\bf a} and {\bf d} are the photoelectron distributions mapped on $k_x$\,-\,$k_y$ plane at $E -E_{\rm F}=$ 0, -0.24 and -0.48~eV (from top to bottom). Shown in {\bf b} and {\bf e} are the dispersions mapped on $k_y=0$ plane. Shown in {\bf c} and {\bf f} are the distributions recorded in the normal-emission configuration. ({\bf g} and {\bf h}) The distribution curves for samples~1 and 2 recorded at some electron energies ({\bf h}) and emission angles ({\bf i}). ({\bf i-l}) Mapping of the fibre-laser ARPES results for sample~1 recorded at $R=0$ ({\bf i}), 2.5 ({\bf j}), 5.0 ({\bf k}) and 7.5$^{\circ}$ ({\bf l}). $D_{\pm}$ in {\bf b}, {\bf e}, and {\bf i-l} are the pair of parabola parameterised in Supplementary Note~2.}
\end{center}
\end{figure*}

In Supplementary Fig.~\ref{figS1}a and b, we show the snapshots of the fibre-laser system contained in a box that has a footprint of $50\times56$~cm$^2$. There are three levels in the container. At the bottom (level~1), current suppliers and temperature controllers for laser diodes are located. At the middle (level~2), a Yb-doped fibre-laser oscillator~\cite{08OptExp_Kobayashi} compacted in the so-called $N$-box of $23\times25$~cm$^2$ footprint~\cite{17OptExpress_Nakamura} is situated. The laser pulses of 1035-nm centre wavelength and $\sim$10~mW generated from the oscillator are sent by an optical fibre to the top (level~3), where the pulses are amplified to $\sim$3~W and up-converted to the fifth harmonics (5.988~eV) by some optical devices fixed on a board of $45\times45$~cm$^2$. Special care was taken for the fibre tip that outputted the amplified $\sim$3-W laser: The space between the tip and subsequent collimating lens was wrapped to avoid dusts from sticking on the tip. 

The oscillator on level~2 weighs 14~kg, and the board-and-device on level~3 weighs 23~kg. In order to isolate the vibration from the container, the oscillator and the board are attached with some rubber feet (AV5/M, Thorlabs). The oscillator was stably mode locked and generating femtosecond pulses for at least three~months. 
This ensured that the profile of the laser such as the photon energy and bandwidth was locked during the measurement. 

In Supplementary Fig.~\ref{figS1}c, we show the beamline that directs the 5.988-eV probe into the vacuum chamber of the spectrometer. Note, the probe can pass through the air, so the beam alignment can be done in atmosphere. We first regulate the power of the probe by the combination of a waveplate\,$\textcircled{\scriptsize5}$  and polariser\,$\textcircled{\scriptsize4}$  (Rochon prism, Kogakugiken), and then control its polarization by a waveplate\,$\textcircled{\scriptsize3}$. Finally, we focus the probe into the chamber by a CaF$_2$ lens\,$\textcircled{\scriptsize2}$ located before a CaF$_2$ window\,$\textcircled{\scriptsize1}$. For the procedure to focus the beam to the focal point of the $e$-lens of the analyser, see Ishida {\it et al.}~\cite{14RSI_TiSi}

Finally, we describe some unique points when aligning the beam for detecting the slow photoelectrons. In general, the beam has to be directed towards the focal point of the $e$-lens. In the DA30-L analyser system of Scienta-Omicron, the focal point is designed to be at $z_{\rm a}-z_{\rm s}=34$~mm from the entrance of the $e$-lens, but this working distance is for high-energy photoelectrons whose trajectory from the sample to the $e$-lens can be regarded as straight. When the photoelectrons are slow, their trajectory can be bent substantially as illustrated in Fig.~\ref{fig4} in the main text, and the focal point can be shifted from the design. In our $e$-lens setup, the sample position for the work-function measurement was approximately 2~mm closer to the $e$-lens than the optimal position for the helium-lamp ARPES measurements around the Fermi cutoff. Practically, we adjusted the aluminium mirror\,$\textcircled{\scriptsize6}$ directing the probe into the beamline (Supplementary Fig.~\ref{figS1}c), so that the surface bands of Au(111) appeared sharp in the photoelectron distribution. 
\\

\subsection*{Supplementary Note~2: Surface bands of Au(111)} 
In Supplementary Fig.~\ref{figS2}, we show the Au(111) surface bands of sample~1 (Supplementary Fig.~\ref{figS2}a-c) and sample~2 (Supplementary Fig.~\ref{figS2}d-f) recorded by using helium-lamp ARPES. Au(111) exhibited spin-split Shockley surface states dispersing around the surface Gamma point ($k_x=k_y=0$) as seen in the distributions in $k_x$\,-\,$k_y$ planes (Supplementary Fig.~\ref{figS2}a and d). In the normal-emission configuration, we observed the spin-split pair of parabolic dispersions (Supplementary Fig.~\ref{figS2}b, c, e and f). The dispersions for both samples 1 and 2 were traced with $D_{\pm}=\beta k_x^2\pm\alpha_{\rm R} k_x-\varepsilon_{\rm D}$ with the parameters $(\beta, \alpha_{\rm R}, \varepsilon_{\rm D}) = (13.8~{\rm eV \AA^2}, 0.32~{\rm eV \AA}, 0.480~{\rm eV})$. The parameters fell into the values found in the literature~\cite{96PRL_LaShell,04PRB_Hoesch,12PRB_BYKim}. The curves of $D_{\pm}$ also agreeably traced the bands seen with fibre-laser ARPES (Supplementary Fig.~\ref{figS2}i-l), which provides another credit that our ARPES setup was operating agreeably for the slow photoelectrons. 

\begin{figure}[htb]
\begin{center}
\includegraphics{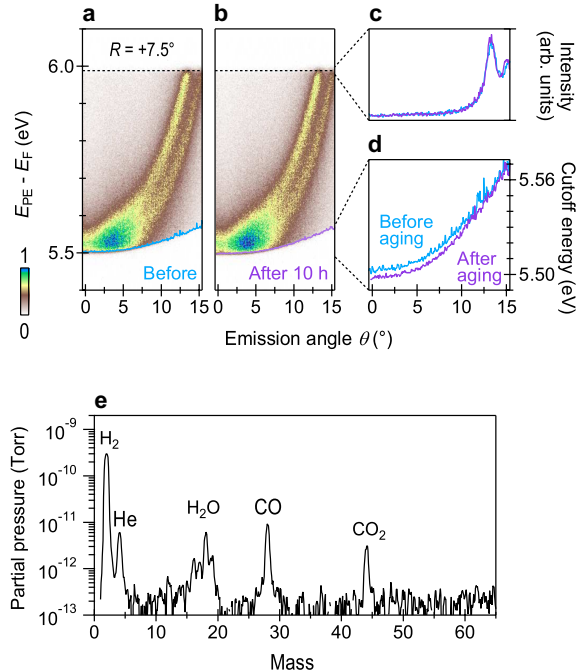}
\caption{\label{figS3}{\bf Aging of the Au(111) surface.} ({\bf a} and {\bf b}) Photoelectron distributions recorded before ({\bf a}) and after ({\bf b}) waiting for 10~hours. The dispersions of the cutoff energy on the slow side are overlaid. ({\bf c}) Distribution curves at the Fermi cutoff. ({\bf d}) Dispersion of the slow-side cutoff. ({\bf e}) Mass spectrum of the residual gas in the vacuum chamber.}
\end{center}
\end{figure}

In Supplementary Fig.~\ref{figS2}g and h, we compare the distribution curves for samples 1 and 2 at some energies and emission angles, respectively. The band of sample~2 was sharper than that of sample~1. This indicated that the surface of sample~2 had a higher quality than that of sample~1 presumably because the pressure during the final annealing at 550~K was at most  $4\times10^{-10}$~Torr for sample~2, whereas it was $7\times10^{-10}$~Torr for sample~1 as described in Methods. 
\\

\subsection*{Supplementary Note~3: Aging of Au(111)} 
Supplementary Fig.~\ref{figS3}a and b respectively show the fibre-laser ARPES results on sample~1 acquired before and after waiting for 10~hours. During the 10~hours, the sample was kept in the ARPES chamber at the pressure of  $3.7\times10^{-11}$~Torr, at $T=30$~K, and without exposure to the 5.988-eV light. The visibility of the distribution curves at the Fermi cutoff (Supplementary Fig.~\ref{figS3}c) hardly changed whereas the cutoff on the slow side shifted downwards for 5.5~meV (Supplementary Fig.~\ref{figS3}d). 

\begin{figure}[htb]
\includegraphics{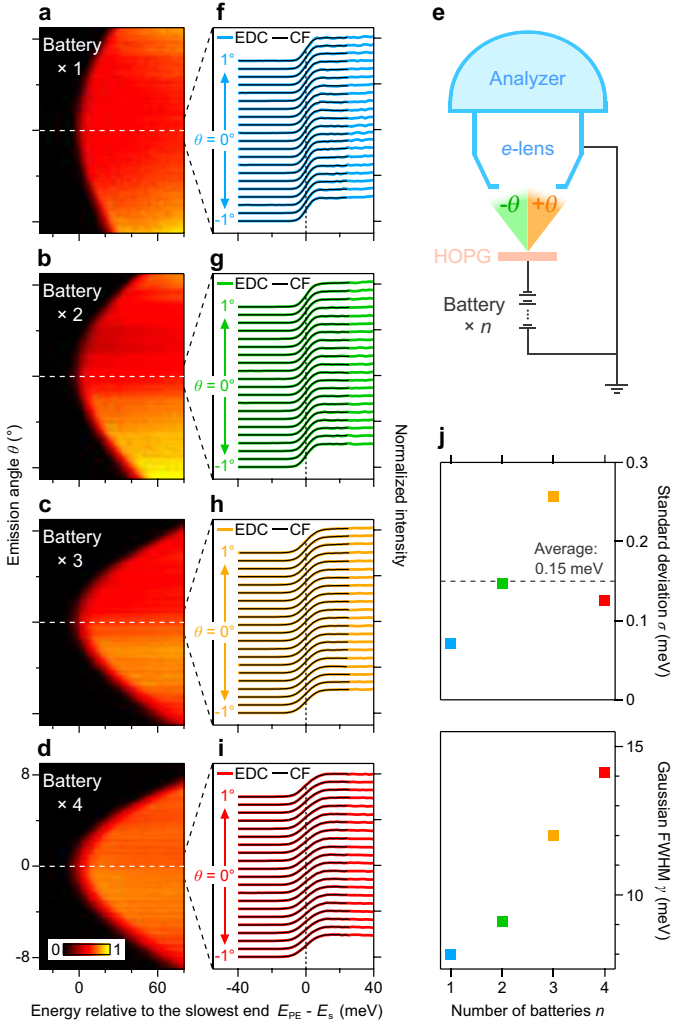}
\caption{\label{figS4}{\bf Bias voltage dependence of the slow-side spectrum.} ({\bf a-d}) ARPES performed on HOPG attatched with batteries up to four ($n=1$ to $4$). ({\bf e}) Schematic illustration of the ARPES setup. ({\bf f-i}) EDCs and CFs for $\theta=\pm1^{\circ}$. ({\bf j}) One standard deviation $\sigma$ (upper panel) of the slowest-end energy extracted from the 11 EDCs about $\pm$0.5$^{\circ}$. The lower panel shows the average of the Gaussian FWHM ($\gamma$) of the 11 CFs.}
\end{figure}

In Supplementary Fig.~\ref{figS3}e, we show the typical mass spectrum of the residual gas in the ARPES chamber. None of the vapor pressure curves for the residual-gas species seen in Supplementary Fig.~\ref{figS3}e, except for the curve of CO$_2$ that had the lowest partial pressure, crosses the level of $3.7\times10^{-11}$~Torr from 30 to 90~K. Therefore, the degassing from the cryostat was negligibly small in the present study performed in the range of 30 to 90~K. Note, at the pressure level of $3.7\times10^{-11}$~Torr, CO , CO$_2$ and H$_2$O respectively vaporize when the temperature exceeds approximately 20, 70 and 130~K. 
\\

\subsection*{Supplementary Note~4: Demonstration of the sub-meV precision when bias voltage is applied} 
In a separate experimental setup, we measured the slow side of the spectrum while appliying bias voltage on the sample. Supplementary Fig.~\ref{figS4}a-d show the ARPES image of highly-oriented pyrolytic graphite (HOPG) attached with batteries up to four ($n=1$ to 4); for the setup, see Supplementary Fig.~\ref{figS4}e. Even though the measurement was performed at the room temperature and at the pressure of $6\times10^{-10}$~Torr, the latter of which is one order of magnitude worse than that during the Au(111) measurement, 
clear parabolic dispersions of the slow-side cutoff were observed. The curvature of the dispersion was strengthened upon increasing the number of the batteries, which is in accordance to the behaviour expected in equation (3) of the main text. 

In order to extract the energy and width of the slowest-end cutoff, we executed the same fitting procedure to that performed on the Au(111) dataset: The energy distribution curves (EDCs) were fitted with the cutoff functions (CFs), see Methods. Supplementary Fig.~\ref{figS4}f-i show the EDCs and CFs within $\theta=\pm1^{\circ}$. The full width at half the maximum of the Gaussian (FWHM; $\gamma$) is the measure of the width of the CF and was as small as 8.0~meV around $\theta=\pm1^{\circ}$ at $n=1$ (lower panel of Supplementary Fig.~\ref{figS4}j). The width of the cutoff did broaden as $n$ was increased (lower panel of Supplementary Fig.~\ref{figS4}j). Nevertheless, the step-edge-type spectral feature was retained and the cutoff energy around $\theta=\pm0.5^{\circ}$, in which there are 11 data points, was leveled with one standard deviation $\sigma$ around 0.15~meV (upper panel of Supplementary Fig.~\ref{figS4}j). Thus, the energy of the slowest end can be read with the sub-meV precision even when the sample is applied with bias voltage. 
\\

\end{document}